\documentclass[11pt]{article}
\usepackage[final]{acl}
\AtBeginDocument{%
  }

\usepackage{times}
\usepackage{latexsym}

\usepackage[T1]{fontenc}
\usepackage[utf8]{inputenc}

\usepackage{microtype}
\usepackage{amsfonts}
\usepackage{multirow}
\usepackage{natbib}
\usepackage{booktabs}
\usepackage{subcaption}
\usepackage[table, dvipsnames]{xcolor}
\usepackage{amsmath} 
\usepackage{array}
\usepackage{tabularx} 
\usepackage{ragged2e} 
\usepackage{siunitx} 
\usepackage{adjustbox} 
\usepackage{todonotes}
\usepackage{enumitem}
\usepackage{mathtools}
\newcolumntype{R}[1]{>{\RaggedRight\arraybackslash}p{#1}}

\usepackage{array,ragged2e,siunitx,makecell,multirow,booktabs,adjustbox,colortbl,xcolor}
\newcolumntype{Q}[1]{>{\RaggedRight\hspace{0pt}\arraybackslash}p{#1}} % Query (we force white per-cell)
\newcolumntype{R}[1]{>{\RaggedRight\hspace{0pt}\arraybackslash}p{#1}} % text, top, left
\newcolumntype{C}[1]{>{\centering\arraybackslash}p{#1}}               % text, top, centered
\newcommand{\topfive}[5]{\makecell{1.\ #1\\2.\ #2\\3.\ #3\\4.\ #4\\5.\ #5}}

\definecolor{groundtruthcolor}{HTML}{FFA500}

\definecolor{Green}{RGB}{34,139,34}
\definecolor{AcademicGreen}{RGB}{34,139,34}
\definecolor{AcademicRed}{RGB}{200,50,50}
\definecolor{Red}{RGB}{200,50,50}

\newcommand{\tighteq}[1]{%
  \begingroup
  \setlength{\abovedisplayskip}{0pt}%
  \setlength{\belowdisplayskip}{0pt}%
  \setlength{\abovedisplayshortskip}{0pt}%
  \setlength{\belowdisplayshortskip}{0pt}%
  #1%
  \endgroup
}
\begin{document}

% \title{\sysname{}: A Simple Yet Efficient Method for Interpretable Dense Retrieval via Embedding Modulation in RAG}

% \title{IMRNNs: Making Dense Passage Retrieval Efficient %Embedding Modulation for 
% and Interpretable %Dense Retrieval
% }

\title{IMRNNs: An Efficient Method for Interpretable Dense Retrieval via Embedding Modulation}

\author{
Yash Saxena$^\dagger$, 
Ankur Padia$^\dagger$, Kalpa Gunaratna$^\ddagger$, Manas Gaur$^\dagger$ \\
$^\dagger$University of Maryland, Baltimore County,
Baltimore, Maryland, USA \\
$^\ddagger$Independent Researcher, 
San Jose, California, USA \\
\texttt{\{ysaxena1,pankur1,manas\}@umbc.edu}, \texttt{gunaratnak@acm.org}
}

\newcommand{\sysname}{{\tt{IMRNNs}}}

\maketitle
\begin{abstract}

Interpretability in black-box dense retrievers remains a central challenge in Retrieval-Augmented Generation (RAG). Understanding how queries and documents semantically interact is critical for diagnosing retrieval behavior and improving model design. However, existing dense retrievers rely on static embeddings for both queries and documents, which obscures this bidirectional relationship. Post-hoc approaches such as re-rankers are computationally expensive, add inference latency, and still fail to reveal the underlying semantic alignment. To address these limitations, we propose Interpretable Modular Retrieval Neural Networks (\sysname{}), a lightweight framework that augments any dense retriever with dynamic, bidirectional modulation at inference time. \sysname{} employs two independent adapters: one conditions document embeddings on the current query, while the other refines the query embedding using corpus-level feedback from initially retrieved documents. This iterative modulation process enables the model to adapt representations dynamically and expose interpretable semantic dependencies between queries and documents. 
Empirically, \sysname{} not only enhances interpretability but also improves retrieval effectiveness. Across the BEIR Benchmark, applying our method to standard dense retrievers yields average gains of +6.35\% in nDCG, +7.14\% in recall, and +7.04\% in MRR over state-of-the-art baselines. These results demonstrate that incorporating interpretability-driven modulation can both explain and enhance retrieval in RAG systems.
\end{abstract}

\section{Introduction}

Retrieval-Augmented Generation (RAG) systems have emerged as a dominant paradigm for grounding large language models (LLMs) in factual, domain-specific knowledge~\cite{lewis2020retrieval, glass2022re2g}. At the heart of these systems lies the \textit{initial retriever}, responsible for selecting candidate documents from massive corpora before downstream re-ranking and generation. This component defines both the \textit{efficiency} and \textit{trustworthiness} of the entire pipeline. However, current dense retrieval methods face a fundamental limitation: they operate with \textit{static embeddings} that encode queries and documents into fixed vector representations, preventing semantic adaptation between them at inference time.

This static nature creates two intertwined problems. First, it limits \textit{retrieval performance}. When query and document embeddings cannot adapt to each other's semantic context, the retriever struggles to capture context-sensitive relevance signals, particularly for complex or ambiguous queries~\cite{li2021more}. Second, and more critically underexplored, it lacks \textit{interpretability}, unlike lexical methods (e.g., \textsc{BM25}) that provide transparent term-matching explanations~\cite{robertson2009probabilistic}, dense retrievers function as black boxes,  preventing users from understanding which aspects of a query influenced retrieval decisions or how specific document features contributed to ranking outcomes~\cite{zhou2024trustworthiness}. This opacity is particularly problematic in high-stakes domains such as healthcare \cite{munnangi2025factehr}, legal research \cite{magesh2025hallucination}, and finance \cite{kim2025optimizing}, where understanding retrieval decisions is essential for trust and accountability.

\begin{figure*}
    \centering
    \includegraphics[width=0.9\linewidth, height=3.2in]{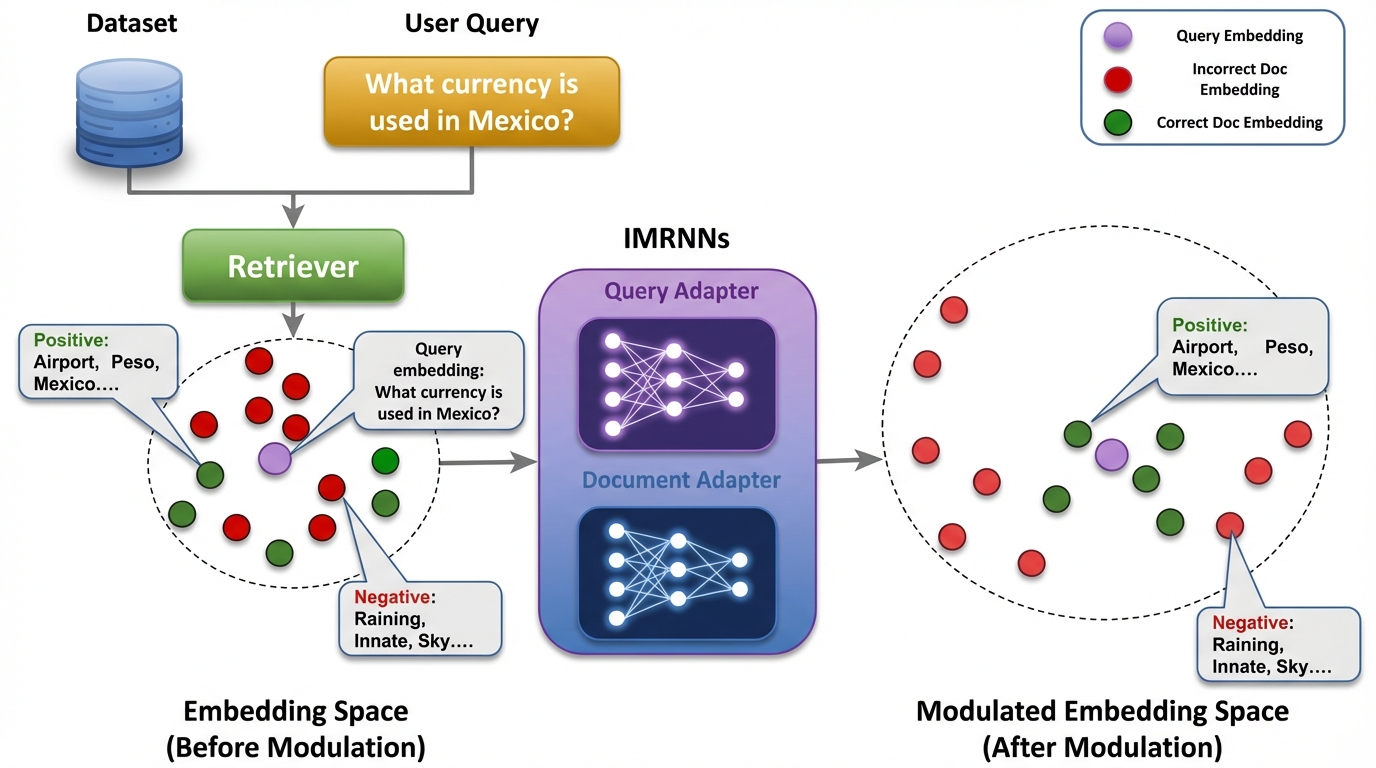}
    \caption{Example illustrating \sysname{}’s modulation mechanism. Starting from a static embedding space produced by the retriever, \sysname{} bidirectionally modulates query and document embeddings to form a modulated embedding space, drawing relevant documents closer to the query while pushing irrelevant ones away. The modulation is interpretable via modulation vectors and their associated key tokens: positive documents align with keywords such as Peso and Mexico, while negative documents align with Raining and Sky.}
    \label{fig:main_example}
\end{figure*}

Recent research has begun addressing retrieval performance through lightweight adaptation mechanisms. \textit{Retrieval adapters}, such as \textsc{search-adaptor}~\cite{yoon2024search} and \textsc{hypencoder}~\cite{killingback2025hypencoder} have applied learned transformations over frozen embeddings, while \textsc{dime}~\cite{campagnano2025unveiling} leverages Matryoshka representations to select important embedding subspaces. However, these methods share a critical gap: \textit{they prioritize computational efficiency and marginal performance gains while providing no semantic-level interpretability}.

\textsc{Search-Adaptor} learns dataset-level correction matrices for queries and documents, adjusting scores uniformly but obscuring which semantic aspects drive changes. \textsc{HypEncoder} generates query-specific MLP scorers to replace cosine similarity, yet these scorer weights do not reveal which query concepts determine relevance. \textsc{DiME} selects top dimensions by $\ell_2$ magnitude (e.g., 256/1024), showing which dimensions matter but not their semantic meaning. All existing adapters operate on dimensions or learned functions without translating to human-understandable concepts.

We introduce \textbf{Interpretable Modular Retrieval Neural Networks (\sysname{})}, the first retrieval adapter providing simultaneous performance gains and multi-level interpretability without external explanation methods. \sysname{} comprises two lightweight MLPs on static embeddings: a \textit{Query Adapter} modulating document embeddings $\mathbf{d}$ conditioned on query $\mathbf{q}$ (pulling relevant documents closer, pushing irrelevant ones away), and a \textit{Document Adapter} aggregating corpus-level signals to adapt the query embedding. \sysname{} uses a bidirectional modulation mechanism to achieve three forms of interpretability \textit{ignored} in prior work:

\begin{itemize}[noitemsep,leftmargin=*]
\item \textit{Structural interpretability}: The modulation mechanism uses explicit affine transformations $\mathbf{W}_q \mathbf{d} + \mathbf{b}_q$ and $\bar{\mathbf{W}}_d \mathbf{q} + \bar{\mathbf{b}}_d$, where $\mathbf{W}_q$ and $\bar{\mathbf{W}}_d$ are learnable weight matrices and $\mathbf{b}_q$ and $\bar{\mathbf{b}}_d$ are learned bias vectors. These parameters are directly observable. Unlike multi-layer neural scorers, users can inspect exactly which mathematical operations transformed the embeddings~\cite{wang2023interpretability, arendt2021crosscheck}.
    
    \item \textit{Attribution-level interpretability}: By computing the difference $\Delta \mathbf{d} = \mathbf{d}_{\text{mod}} - \mathbf{d}_{\text{orig}}$ and $\Delta \mathbf{q} = \mathbf{q}_{\text{mod}} - \mathbf{q}_{\text{orig}}$, where $\mathbf{d}_{\text{mod}}$ and $\mathbf{q}_{\text{mod}}$ are the modulated embeddings, we obtain the exact change vector induced by modulation. This reveals precisely which embedding dimensions increased or decreased, and by how much, enabling dimension-level attribution of retrieval decisions~\cite{zhou2022feature, zhang2022locally, calderon2025behalf}.

\item \textit{Semantic-level interpretability}: We back-project the change vectors $\Delta \mathbf{d}$ and $\Delta \mathbf{q}$ from the adapter's working space to the original encoder's token embedding space using the Moore-Penrose pseudoinverse~\cite{barata2012moore}. Let $P^{+}$ denote this pseudoinverse. By computing cosine similarity between the back-projected vector and every token embedding, we identify tokens whose semantics align with the modulation direction, revealing \textit{what semantic concepts} drove the retrieval decision~\cite{rajagopal2021selfexplain, sajjad2022neuron}.
\end{itemize}

Crucially, the identified tokens are not post-hoc explanations but directly correspond to the mathematical transformations that changed rankings. This contrasts with prior interpretability approaches~\cite{yuksel2025interpretability,llordes2023explain} that analyze model behavior after the fact rather than exposing the mechanism itself. \sysname{} adapts per-query while preserving cosine similarity's efficiency. \autoref{fig:main_example} shows the complete end-to-end workflow.

\noindent \textbf{Contributions.} Our work makes the following contributions:
\begin{itemize}[noitemsep,leftmargin=*]
    \item We propose \sysname{}, the \textbf{first retrieval adapter} with structural, attribution-level, and semantic-level interpretability for dense retrieval systems.
    \item We design a \textbf{bidirectional per-query modulation mechanism} enabling semantic alignment without re-encoding or expensive cross-attention.
    \item We develop a \textbf{token-level attribution method} using Moore-Penrose back-projection that causally links embedding modulations to human-interpretable keywords.
    \item We demonstrate \textbf{significant performance gains} across seven diverse benchmarks with minimal computational overhead.
\end{itemize}

% Revised Related Work Section for IMRNNs

\section{Related Work}
\label{sec:related}

We organize related work into three categories: dense retrieval foundations, adapter mechanisms for performance enhancement, and the interpretability gap we address.

\noindent \textbf{Dense Retrieval Methods.} RAG systems employ a two-stage pipeline: an initial retriever selects candidates efficiently, followed by a reranker. Initial retrievers include lexical methods like \textsc{bm25}~\citep{10.1561/1500000019} with transparent term-matching, dense bi-encoders like \textsc{DPR}~\citep{karpukhin-etal-2020-dense} and Contriever~\citep{izacard2022unsupervised} capturing semantic similarity through learned embeddings, and hybrid approaches like \textsc{splade}~\citep{formal-2021-splade} combining contextual encoders with sparse representations. We focus on dense bi-encoders because they dominate modern RAG systems yet operate as black boxes.

\noindent \textbf{Retrieval Adapters.} Recent work has introduced lightweight adapters that enhance dense retrievers without retraining. \textit{Architecture-modifying adapters}~\citep{10.1145/3627673.3680095,zeighami2025nudge,ding2023parameter} update internal components, requiring white-box access. \textit{Embedding-space adapters} operate on frozen encoder outputs: \textsc{search-adaptor}~\citep{yoon-etal-2024-search}, \textsc{dime}~\citep{10.1145/3726302.3730318}, and \textsc{hypencoder}~\citep{10.1145/3726302.3729983} represent state-of-the-art embedding-space adaptation. \textit{Embedding compression methods}~\citep{liu-etal-2022-dimension,ma-etal-2021-simple} reduce dimensionality but do not explain how transformations affect query-document interactions. \textit{Parameter-efficient fine-tuning approaches} modify encoder parameters minimally: \textsc{tart}~\citep{asai-etal-2023-task} and \textsc{instructor}~\citep{su-etal-2023-one} inject task-specific instructions; LoRA~\citep{hu2022lora} and IA3~\citep{10.5555/3600270.3600412} update low-rank subspaces; \textsc{promptagator}~\citep{dai2023promptagator} synthesizes training data via LLMs. We exclude these from our experimental comparison because they require encoder fine-tuning (violating the frozen-encoder constraint that enables \sysname{} plug-and-play deployment) and still provide no interpretability mechanisms. \textsc{Adapted Dense Retrieval}~\citep{khatry2023augmentedembeddingscustomretrievals} learns low-rank residuals for heterogeneous retrieval settings. Adapters have also been explored for sparse retrievers and rerankers~\citep{hu-etal-2023-llm}, which fall outside our focus on dense initial retrieval. While these methods improve retrieval accuracy through various adaptation mechanisms, \textbf{none provide interpretability}: users cannot determine which semantic features drove retrieval decisions or how queries and documents are semantically aligned during ranking.

% \noindent \textbf{Interpretability in Information Retrieval.} Interpretability has been explored in other RAG components, dense retrievers remain largely unaddressed. Existing approaches have critical limitations:

\noindent \textbf{Interpretability in Dense retrievers.}  Interpretability for dense retrievers remain largely unaddressed, and existing approaches have critical limitations

\begin{itemize}[noitemsep,leftmargin=*]

    \item \textit{Surrogate approximations}~\citep{10.1145/3539618.3591982} fit sparse models to approximate dense rankings, but these post-hoc explanations may not faithfully represent the actual decision process, and approximation quality degrades as the sparse-dense gap widens.

    \item \textit{Gradient-based attribution}~\citep{yuksel2025interpretabilityanalysisdomainadapted} identifies high-gradient tokens during training but does not reveal semantic concepts emphasized during inference, and gradient explanations can be unstable~\citep{adebayo2018sanity}.

    \item \textit{Concept mapping}~\citep{kang-etal-2025-interpret} aligns embedding dimensions with human-interpretable descriptors via sparse probing, but requires additional annotation and does not explain query-document interactions during retrieval.

\end{itemize}

% Problem Formulation Section for IMRNNs

\section{Problem Formulation}
\label{sec:problem}

% This section formally defines the dense retrieval task and introduces the \sysname{} architecture. 
%The complete implementation details are provided in Section~\ref{sec:approach}, training procedures in Section~\ref{sec:experiments}, and empirical validation in Section~\ref{sec:results}.

\paragraph{}
\noindent \textbf{Task Definition:} Given a user query $q$ and a document corpus $\mathcal{D} = \{d_1, d_2, \ldots, d_N\}$, the goal of the dense retriever is to rank all documents in $\mathcal{D}$ by their relevance to $q$. 
% Let $\mathcal{Q}$ denote the set of all queries. \todo{Where is the $\mathcal{Q}$ used in the text ? If used, i could not find, make a one sentence description here.} 
A dense retrieval system consists of two components:(a) A \textbf{base encoder} $f_\theta: \mathcal{V} \rightarrow \mathbb{R}^n$ that maps text sequences (queries or documents) to fixed-dimensional embeddings, where $\mathcal{V}$ is the space of all possible text sequences and $n$ is the embedding dimension. (b) A \textbf{similarity function} $s: \mathbb{R}^n \times \mathbb{R}^n \rightarrow \mathbb{R}$ that computes relevance scores between query and document embeddings. Standard dense retrievers compute $ \text{score}(q, d_i) = \cos\big(f_\theta(q), f_\theta(d_i)\big)$, 
where $\cos(\cdot, \cdot)$ denotes cosine similarity. Documents are then ranked in descending order of these scores.
% \todo{Ankur: Is this limitation of your work is it "Previous work limitation" or "Conventional approach"}

\paragraph{}
\noindent \textbf{Current Limitation With Static Embeddings}: The embeddings $f_\theta(q)$ and $f_\theta(d_i)$ are \textit{static}, computed independently and fixed after encoding. This means a document receives the same embedding regardless of which query it is being matched against, preventing the retriever from dynamically emphasizing query-relevant aspects of each document. Additionally, the system lacks interpretability: users cannot see which semantic dimensions drive the similarity score for a specific query-document pair.

\subsection{Our approach: \textbf{\sysname{}}}

\sysname{} address this limitation by introducing \textit{dynamic modulation} on top of static embeddings from a frozen base encoder. %\textcolor{red}{The architecture consists of three components (detailed architectural specifications are provided in Section~\ref{sec:approach}):}

\noindent \textbf{1. Dimension Reduction via Projection.} To enable efficient learning and generalization, we project the high-dimensional base embeddings into a lower-dimensional working space. Let $\mathbf{P} \in \mathbb{R}^{m \times n}$ be a fixed linear projection matrix, where $m < n$ (we use $m=256$ for $n=1024$ in our experiments). For query $q$ and document $d_i$, we compute:
\begin{align*}
    \mathbf{q}_{\text{orig}} &= f_\theta(q) \in \mathbb{R}^n, \quad \mathbf{q}_{\text{proj}} = \mathbf{P} \mathbf{q}_{\text{orig}} \in \mathbb{R}^m \\
    \mathbf{d}_{\text{orig}}^{(i)} &= f_\theta(d_i) \in \mathbb{R}^n, \quad \mathbf{d}_{\text{proj}}^{(i)} = \mathbf{P} \mathbf{d}_{\text{orig}}^{(i)} \in \mathbb{R}^m
\end{align*}

We optimize $\mathbf{P}$ jointly with the adapters (described next), while the encoder $f_\theta$ remains frozen.

\paragraph{2. Query Adapter} ($\mathcal{A}_q$) is a lightweight neural network that uses projected query embedding to  produce a weight matrix and bias vector to modulate all document embeddings:
$ \mathcal{A}_q(\mathbf{q}_{\text{proj}}) = \big(\mathbf{W}_q \in \mathbb{R}^{m \times m}, \mathbf{b}_q \in \mathbb{R}^m\big) $. The modulated document embedding for document $d_i$ is then computed as:
$ \mathbf{d}_{\text{mod}}^{(i)} = \mathbf{W}_q \cdot \mathbf{d}_{\text{proj}}^{(i)} + \mathbf{b}_q $. This transformation allows the query to \textit{pull} semantically relevant documents closer in the embedding space and \textit{push} irrelevant documents farther away, adapting document representations to the specific query context.

\paragraph{3. Document Adapter} ($\mathcal{A}_d$) is a separate second lightweight neural network that processes each document embedding independently to produce document-specific transformations:
$ \mathcal{A}_d(\mathbf{d}_{\text{proj}}^{(i)}) = \big(\mathbf{W}_d^{(i)} \in \mathbb{R}^{m \times m}, \mathbf{b}_d^{(i)} \in \mathbb{R}^m\big)$. These transformations are aggregated across all documents to create a corpus-level adaptation signal:
\tighteq{
\[
\bar{\mathbf{W}}_d = \frac{1}{N} \sum_{i=1}^{N} \mathbf{W}_d^{(i)}, \quad \bar{\mathbf{b}}_d = \frac{1}{N} \sum_{i=1}^{N} \mathbf{b}_d^{(i)} \]}
The modulated query embedding is then computed as:
$\mathbf{q}_{\text{mod}} = \bar{\mathbf{W}}_d \cdot \mathbf{q}_{\text{proj}} + \bar{\mathbf{b}}_d$. This enables the query embedding to adapt to the characteristics of the document corpus, thereby aligning with the vocabulary and semantic space of the available documents.

\paragraph{4. Scoring and Ranking.} After bidirectional modulation, the final relevance score between query $q$ and document $d_i$ is computed using cosine similarity:
\tighteq{
\[
    \text{score}_{\text{IMRNNs}}(q, d_i) = \cos\big(\mathbf{q}_{\text{mod}}, \mathbf{d}_{\text{mod}}^{(i)}\big)
\]}

Layer normalization is applied to both $\mathbf{q}_{\text{mod}}$ and $\mathbf{d}_{\text{mod}}^{(i)}$ before computing cosine similarity to ensure stable gradients and bounded scores.

\subsection{Training}
\sysname{} are trained using a margin-based ranking loss over query-document pairs. For each query $q$ in a training batch $\mathcal{B}$, we sample one relevant document $d^+$ and one irrelevant document $d^-$ with BM25 for a hard negative example. The loss function is:
\begin{equation*}
\begin{split}
    \mathcal{L} = \frac{1}{|\mathcal{B}|} \sum_{(q, d^+, d^-) \in \mathcal{B}} \max\Big\{0, \gamma - \cos(\mathbf{q}_{\text{mod}}, \mathbf{d}_{\text{mod}}^+) \\
    \quad + \cos(\mathbf{q}_{\text{mod}}, \mathbf{d}_{\text{mod}}^-)\Big\}
\end{split}
\end{equation*}
where $\gamma > 0$ is a margin hyperparameter that enforces a minimum separation between relevant and irrelevant pairs in the modulated embedding space.

During training, only the parameters of $\mathcal{A}_q$ and $\mathcal{A}_d$ are updated while the base LLM encoder (such as e5-large, MiniLM, and BGE) \citep{wang2024textembeddingsweaklysupervisedcontrastive},  $f_\theta$ remain frozen. We optimize using the Adam optimizer with weight decay regularization to prevent overfitting. To bound computational cost and memory requirements, training operates on a subset of top-$k$ BM25-retrieved candidates per query rather than the entire corpus. Early stopping is applied based on validation set performance to determine the optimal number of training epochs. Additional hyperparameter settings and implementation details are provided in Section~\ref{sec:experiments}.

\subsection{Inference Workflow}

At inference time, \sysname{} operate as follows: 

\begin{enumerate}[noitemsep,leftmargin=*]
    \item For a new query $q$, compute $\mathbf{q}_{\text{orig}} = f_\theta(q)$ and $\mathbf{q}_{\text{proj}} = \mathbf{P} \mathbf{q}_{\text{orig}}$. For all documents in the corpus (pre-computed offline), obtain $\{\mathbf{d}_{\text{proj}}^{(i)}\}_{i=1}^N$.
    
    \item Pass $\mathbf{q}_{\text{proj}}$ through $\mathcal{A}_q$ to obtain $(\mathbf{W}_q, \mathbf{b}_q)$, then compute modulated document embeddings:
    \begin{equation*}
        \mathbf{d}_{\text{mod}}^{(i)} = \mathbf{W}_q \cdot \mathbf{d}_{\text{proj}}^{(i)} + \mathbf{b}_q \quad \text{for } i = 1, \ldots, N
    \end{equation*}
    
    \item  Pass each $\mathbf{d}_{\text{proj}}^{(i)}$ through $\mathcal{A}_d$ to obtain $\{(\mathbf{W}_d^{(i)}, \mathbf{b}_d^{(i)})\}_{i=1}^N$, then aggregate and modulate the query:
    \tighteq{
    \[
    \mathbf{q}_{\text{mod}} = \bar{\mathbf{W}}_d \cdot \mathbf{q}_{\text{proj}} + \bar{\mathbf{b}}_d
        %\mathbf{q}_{\text{mod}} = \left(\frac{1}{N} \sum_{i=1}^{N} \mathbf{W}_d^{(i)}\right) \mathbf{q}_{\text{proj}} + \frac{1}{N} \sum_{i=1}^{N} \mathbf{b}_d^{(i)}
    \]}
    
    \item For each document $d_i$, compute $\text{score}_{\text{IMRNNs}}(q, d_i) = \cos(\mathbf{q}_{\text{mod}}, \mathbf{d}_{\text{mod}}^{(i)})$ and rank documents in descending order of score.
\end{enumerate}

\noindent \textit{Computational Efficiency:} The adapter networks $\mathcal{A}_q$ and $\mathcal{A}_d$ are lightweight (2-layer MLPs), adding minimal overhead. Document projections $\mathbf{d}_{\text{proj}}^{(i)}$ can be pre-computed offline and cached, so adapter forward passes dominate online cost per query and scale linearly with corpus size.

\subsection{Interpretability Mechanism}

The modulation framework enables analyzing the change induced in the original embedding space for direct interpretability. For any query-document pair $(q, d_i)$, we compute the modulation vectors:
\tighteq{
\begin{align*}
    \Delta \mathbf{q} &= \mathbf{q}_{\text{mod}} - \mathbf{q}_{\text{proj}} \\
    \Delta \mathbf{d}^{(i)} &= \mathbf{d}_{\text{mod}}^{(i)} - \mathbf{d}_{\text{proj}}^{(i)}
\end{align*}
}
The change in retrieval score is measured as:
\tighteq{
\begin{equation*}
    \Delta \text{similarity} = \cos(\mathbf{q}_{\text{mod}}, \mathbf{d}_{\text{mod}}^{(i)}) - \cos(\mathbf{q}_{\text{proj}}, \mathbf{d}_{\text{proj}}^{(i)})
\end{equation*}
}

Positive $\Delta$ similarity indicates the modulation pulled the query-document pair closer (increasing relevance), while negative values indicate they were pushed apart (decreasing relevance). To interpret these vectors in terms of semantic concepts, we back-project them to the original encoder's embedding space using the Moore-Penrose pseudoinverse $\mathbf{P}^+ = \mathbf{P}^{\top}(\mathbf{P}\mathbf{P}^{\top})^{-1}$ as follows:
\tighteq{
\begin{align*}
    \Delta \mathbf{q}_{\text{orig}} &= \mathbf{P}^+ \Delta \mathbf{q} \\
    \Delta \mathbf{d}_{\text{orig}}^{(i)} &= \mathbf{P}^+ \Delta \mathbf{d}^{(i)}.
\end{align*} }
In general, with $\mathbf{P} = \mathbf{U}\boldsymbol{\Sigma}\mathbf{V}^{\top}$ (SVD), we can say that  $\mathbf{P}^{+} = \mathbf{V}\boldsymbol{\Sigma}^{+}\mathbf{U}^{\top}$.
%\({P}^{+}={V}\boldsymbol{\Sigma}^{+}{U}^{\top}\).

Let $\mathbf{E} \in \mathbb{R}^{|\mathcal{V}| \times n}$ denote the encoder's token embedding table, where each row $\mathbf{e}_t$ is the embedding of token $t \in \mathcal{V}$. We compute cosine similarity between the back-projected modulation vector and each token embedding:
\begin{equation*}
    \text{score}_t = \cos(\Delta \mathbf{q}_{\text{orig}}, \mathbf{e}_t) = \frac{\Delta \mathbf{q}_{\text{orig}}^{\top} \mathbf{e}_t}{\|\Delta \mathbf{q}_{\text{orig}}\|_2 \|\mathbf{e}_t\|_2}
\end{equation*}

Tokens with high positive scores indicate concepts that the modulation emphasized (pulling the embedding toward), while tokens with high negative scores indicate concepts that were de-emphasized (pushing away). By ranking tokens by $|\text{score}_t|$ and examining the top-ranked tokens, we obtain human-interpretable explanations of what semantic features drove the retrieval decision.
\section{Experimental Setup}
\label{sec:experiments}

We evaluate \sysname{} on two complementary tasks across diverse datasets, comparing against state-of-the-art retrieval adapter baselines. 

\paragraph{\textbf{Retrieval Task}.} In the retrieval task, we determine the effectiveness of \sysname{} by comparing document retrieval accuracy with the base dense retriever and recent popular competing adaptation methods. Performance gain is evaluated on BEIR's held-out test sets using MRR, nDCG, and Recall.

\paragraph{\textbf{Interpretability Task}.} We qualitatively examine the semantic transparency of \sysname{}' modulation mechanism. For selected queries, we analyze modulation vectors for both relevant and irrelevant documents, computing the $\Delta$ similarity (the change from the original to the modulated scores) and back-projecting the modulation vectors to identify top-ranked tokens using the Moore-Penrose pseudoinverse (Section~\ref{sec:problem}). We validate interpretability by examining whether the identified keywords align with document relevance and whether changes in ranking correlate with meaningful semantic shifts.
\paragraph{Underlying Base Retrievers}: We consider e5-large-v2, MiniLM-V6-L2 and BGE-Large-en as base retrievers and use it as baselines. 

\paragraph{Dataset Selection Rationale.}  We selected seven datasets from the BEIR benchmark (Apache License 2.0) to ensure broad coverage in evaluating \sysname{} across multiple dimensions. MS MARCO tests scalability with large-scale web search (8.8M documents). Natural Questions and HotpotQA use Wikipedia, where HotpotQA requires multiple passages for multi-hop reasoning to synthesize answers, testing whether modulation connects semantically related but lexically distinct evidence. SciFact and TREC-COVID evaluate domain-specific terminology and precise semantic matching in scientific/biomedical retrieval. FiQA-2018 tests adaptation to financial jargon and numerical reasoning. Webis-Touché 2020 involves argumentation retrieval and depends on identifying viewpoints rather than topical overlap. Together, these span corpus sizes from 5K to 8.8M documents, single-hop to multi-hop reasoning, and general to highly specialized domains. More details are available in \autoref{sec:Dataset_selection}.
%Details regarding dataset selection and the corresponding dataset statistics are in Table~\ref{app:dataset-details}, presented in \autoref{sec:Dataset_selection}.

\begin{table*}[t]
\centering
\footnotesize
\caption{\textbf{Qualitative Examples of Modulation Effects on Retrieval Performance.} For each query, we show a relevant (green) and irrelevant (red) document pair. Modulation modestly reduces similarity for relevant documents ($\Delta$ Sim. $\approx -0.06$) while drastically reducing it for irrelevant ones ($\Delta$ Sim. $\approx -0.30$), improving discrimination. This enables relevant documents to rise in rankings (e.g., entering the top-5 as shown in the rightmost columns). Keywords extracted from queries and documents aid interpretability. \colorbox{green!15}{Green} indicates relevant; \colorbox{red!10}{red} indicates irrelevant.}
\label{tab:qualitative_analysis}
\resizebox{\textwidth}{!}{%
\begin{tabular}{@{}%
  Q{2.05cm} % Query (white per-cell via \cellcolor{white})
  C{1.00cm} % Type
  R{2.55cm} % Document
  S[table-format=1.2] % Orig
  S[table-format=1.2] % Mod
  S[table-format=-1.2] % Δ
  R{1.35cm} % Query KW
  R{1.55cm} % Doc KW
  R{1.20cm} % Orig Top-5?  (top-left; compact via \makecell)
  R{1.20cm} % Mod  Top-5?  (top-left; compact via \makecell)
@{}}
\toprule
\textbf{Query} & \textbf{Document Type} & \textbf{Document Text}
& \textbf{Orig. Sim.} & \textbf{Mod. Sim.} & \textbf{$\Delta$ Sim.}
& \textbf{Query Keywords} & \textbf{Document Keywords}
& \textbf{Orig Top-5 Doc. (Before)} & \textbf{Mod Top-5 Doc. (After)} \\
\midrule

% ---------- Block 2 ----------
\rowcolor{green!15}
\multirow{2}{=}{\cellcolor{white}\RaggedRight What currency is used in Mexico?}
& Relevant
& Mexico, Peso. The Mexican Peso is the currency of Mexico. Our currency rankings show that the most popular Mexico Peso exchange \dots
& 0.94 & 0.87 & -0.06
& \RaggedRight Mexican, Ridges
& \RaggedRight Airport, Peso, Mexico
& \topfive{No}{\cellcolor{green!30}\textbf{Yes}}{No}{No}{No}
& \topfive{\cellcolor{green!30}\textbf{Yes}}{No}{No}{No}{No} \\
\rowcolor{red!10}\cellcolor{white}
& Irrelevant
& \textit{All you have to do is tune to the right channel or visit any number of weather and news Web sites and \dots}
& 0.39 & 0.05 & -0.34
& \RaggedRight Mexican, Ridges
& \RaggedRight Raining, Innate, Sky
& -- & -- \\
\midrule

\rowcolor{green!15}
\multirow{2}{=}{\cellcolor{white}\RaggedRight What county is incline village nv?}
& Relevant
& (Redirected from Incline Village\textendash Crystal Bay, Nevada) Incline Village is a census\textendash designated place in Washoe County, Nevada on the north shore \dots
& 0.91 & 0.85 & -0.06
& \RaggedRight Incline, Rupert
& \RaggedRight Incline, Hilly, Nevada
& \topfive{No}{No}{No}{No}{No}
& \topfive{No}{No}{No}{No}{\cellcolor{green!30}\textbf{Yes}} \\
\rowcolor{red!10}\cellcolor{white}
& Irrelevant
& \textit{Public Meetings: The Alva Regional Airport Commission will \dots}
& 0.41 & 0.11 & -0.29
& \RaggedRight Incline, Rupert
& \RaggedRight Airport, Regional, Sanctioned
& -- & -- \\

% ---------- Block 3 ----------
% \rowcolor{green!15}
% \multirow{2}{=}{\cellcolor{white}\RaggedRight Do diabetic patients with acute coronary syndrome experience decreased short-term and long-term risk for bleeding events?}
% & Relevant
% & \RaggedRight BACKGROUND Diabetes mellitus is a major risk factor for adverse outcomes after acute coronary syndromes (ACS). Because \dots
% & 0.93 & 0.84 & -0.08
% & \RaggedRight Coronary, Risk, Bleeding
% & \RaggedRight Diabetes, Increased, Senses
% & \topfive{No}{No}{No}{No}{No}
% & \topfive{No}{No}{\cellcolor{green!30}\textbf{Yes}}{No}{No} \\
% \cellcolor{white}
% \rowcolor{red!10}
% & Irrelevant
% & \textit{Global, regional, and national comparative risk assessment of 79 behavioural, environmental \dots}
% & 0.53 & 0.19 & -0.34
% & \RaggedRight Coronary, Risk, Bleeding
% & \RaggedRight Behavioural, Metabolism
% & -- & -- \\
\bottomrule
\end{tabular}}
\end{table*}

\paragraph{Baselines.} We compare three adapter families: 
\begin{itemize}[noitemsep,leftmargin=*]
\item \textbf{\textsc{dime} variants}: DIME variants represent query-only modulation via dimension selection. Select embedding subspaces by ranking dimensions via $\ell_2$ magnitude. We evaluate 20\%, 40\%, 60\%, and 80\% dimensionality reduction to test whether simple dimension selection competes with learned modulation.
\item \textbf{\textsc{search-adaptor}}: Learns dataset-specific residual transformations applied globally to all queries and documents. This evaluates static, dataset-level adaptation.
% suffices versus \sysname{}' per-query modulation. 
\item \textbf{\textsc{hypencoder} variants}: HYPENCODER variants represent query-only modulation via neural scoring. Generates query-specific MLPs to score documents, replacing cosine similarity with learned neural functions. We evaluate two, four, six, and eight hidden layers configurations to consider increasing expressive scoring functions. 
% test whether more expressive scoring functions outperform \sysname{}' simpler modulation while maintaining interpretability. 
\item \textbf{\sysname{}} is our approach with query and document modulation.
\end{itemize}
All baselines operate on identical frozen encoder embeddings and use the same data splits. Implementation details and the computational efficiency comparison are provided in \autoref{app:impl-details} and \autoref{sec:compute-time} respectively. Finally, we used Moore-Penrose pseudoinverse to map modulation vectors back to actual vocabulary tokens. 

\section{Discussion on Interpretability}

\textbf{Interpretability Task.} Table~\ref{tab:qualitative_analysis} shows that \sysname{} explain retrieval decisions through influential keywords that directly cause score changes. Consider the query \texttt{What currency is used in Mexico?} and the extracted keywords \{\texttt{peso}, \texttt{Mexico}\}. As shown in the Table \ref{tab:qualitative_analysis} the similarity of the relevant document after modulation drops slightly from 0.94 to 0.87 ($\Delta$ = -0.07) as the modulation strengthened the extracted keywords in the embedding thereby refining the match. In contrast, for the same query with keywords \{\texttt{raining}, \texttt{sky}\}, the score for an irrelevant document scores plummets significantly from 0.39 to 0.05 ($\Delta$ = -0.34) as the modulation identified off-topic concepts and pushed the document away. Here the keywords explain a shift in the embeddings, bringing rel relevant concepts closer and pushing away from less relevant ones.

\begin{table*}[tb]
\centering
\footnotesize
\caption{\textbf{Results on Open-Domain Datasets (MS MARCO, Natural Questions, and HotpotQA).}}
\label{tab:ms-nq-hotpot}
\resizebox{\textwidth}{!}{%
\begin{tabular}{l|ccc|ccc|ccc|ccc}
\toprule
\textbf{Methods} &
\multicolumn{3}{c|}{\textbf{MS MARCO}} &
\multicolumn{3}{c|}{\textbf{Natural Questions}} &
\multicolumn{3}{c|}{\textbf{HotpotQA}} &
\multicolumn{3}{c}{\textbf{Average}} \\
& nDCG & R & MRR & nDCG & R & MRR & nDCG & R & MRR & nDCG & R & MRR \\
\midrule
e5\mbox{-}large\mbox{-}v2
& \cellcolor{red!10}0.85 & \cellcolor{red!10}0.97 & \cellcolor{red!10}0.81
& \cellcolor{red!10}0.72 & \cellcolor{red!10}0.90 & \cellcolor{red!10}0.68
& \cellcolor{red!10}0.54 & \cellcolor{red!10}0.61 & \cellcolor{red!10}0.65
& \cellcolor{red!10}0.70 & \cellcolor{red!10}0.83 & \cellcolor{red!10}0.71 \\
MiniLM
& \cellcolor{red!10}0.91 & \cellcolor{red!10}0.97 & \cellcolor{red!10}0.90
& \cellcolor{red!10}0.89 & \cellcolor{red!10}0.68 & \cellcolor{red!10}0.73
& \cellcolor{red!10}0.61 & \cellcolor{red!10}0.64 & \cellcolor{red!10}0.53
& \cellcolor{red!10}0.80 & \cellcolor{red!10}0.76 & \cellcolor{red!10}0.72 \\
BGE
& \cellcolor{red!10}0.88 & \cellcolor{red!10}0.97 & \cellcolor{red!10}0.89
& \cellcolor{red!10}0.73 & \cellcolor{red!10}0.90 & \cellcolor{red!10}0.69
& \cellcolor{red!10}0.61 & \cellcolor{red!10}0.64 & \cellcolor{red!10}0.54
& \cellcolor{red!10}0.74 & \cellcolor{red!10}0.84 & \cellcolor{red!10}0.71 \\
DIME 20\%
& \cellcolor{red!10}0.85 & \cellcolor{red!10}0.97 & \cellcolor{red!10}0.81
& \cellcolor{red!10}0.72 & \cellcolor{red!10}0.89 & \cellcolor{red!10}0.68
& \cellcolor{red!10}0.53 & \cellcolor{red!10}0.60 & \cellcolor{red!10}0.64
& \cellcolor{red!10}0.70 & \cellcolor{red!10}0.82 & \cellcolor{red!10}0.71 \\
DIME 40\%
& \cellcolor{red!10}0.84 & \cellcolor{red!10}0.97 & \cellcolor{red!10}0.80
& \cellcolor{red!10}0.72 & \cellcolor{red!10}0.90 & \cellcolor{red!10}0.67
& \cellcolor{red!10}0.51 & \cellcolor{red!10}0.58 & \cellcolor{red!10}0.62
& \cellcolor{red!10}0.69 & \cellcolor{red!10}0.82 & \cellcolor{red!10}0.70 \\
DIME 60\%
& \cellcolor{red!10}0.83 & \cellcolor{red!10}0.96 & \cellcolor{red!10}0.79
& \cellcolor{red!10}0.70 & \cellcolor{red!10}0.88 & \cellcolor{red!10}0.66
& \cellcolor{red!10}0.47 & \cellcolor{red!10}0.54 & \cellcolor{red!10}0.57
& \cellcolor{red!10}0.67 & \cellcolor{red!10}0.79 & \cellcolor{red!10}0.67 \\
DIME 80\%
& \cellcolor{red!10}0.80 & \cellcolor{red!10}0.93 & \cellcolor{red!10}0.75
& \cellcolor{red!10}0.64 & \cellcolor{red!10}0.83 & \cellcolor{red!10}0.60
& \cellcolor{red!10}0.37 & \cellcolor{red!10}0.42 & \cellcolor{red!10}0.45
& \cellcolor{red!10}0.60 & \cellcolor{red!10}0.73 & \cellcolor{red!10}0.60 \\
SearchAd
& \cellcolor{red!10}0.84 & \cellcolor{red!10}0.96 & \cellcolor{red!10}0.80
& \cellcolor{red!10}0.71 & \cellcolor{red!10}0.89 & \cellcolor{red!10}0.67
& \cellcolor{red!10}0.37 & \cellcolor{red!10}0.43 & \cellcolor{red!10}0.46
& \cellcolor{red!10}0.64 & \cellcolor{red!10}0.76 & \cellcolor{red!10}0.64 \\
Hyp (2)
& \cellcolor{red!10}0.68 & \cellcolor{red!10}0.74 & \cellcolor{red!10}0.71
& \cellcolor{red!10}0.43 & \cellcolor{red!10}0.60 & \cellcolor{red!10}0.40
& \cellcolor{red!10}0.29 & \cellcolor{red!10}0.25 & \cellcolor{red!10}0.27
& \cellcolor{red!10}0.47 & \cellcolor{red!10}0.53 & \cellcolor{red!10}0.46 \\
Hyp (4)
& \cellcolor{red!10}0.69 & \cellcolor{red!10}0.74 & \cellcolor{red!10}0.70
& \cellcolor{red!10}0.44 & \cellcolor{red!10}0.60 & \cellcolor{red!10}0.40
& \cellcolor{red!10}0.29 & \cellcolor{red!10}0.26 & \cellcolor{red!10}0.27
& \cellcolor{red!10}0.47 & \cellcolor{red!10}0.53 & \cellcolor{red!10}0.46 \\
Hyp (6)
& \cellcolor{red!10}0.71 & \cellcolor{red!10}0.75 & \cellcolor{red!10}0.71
& \cellcolor{red!10}0.44 & \cellcolor{red!10}0.61 & \cellcolor{red!10}0.40
& \cellcolor{red!10}0.30 & \cellcolor{red!10}0.26 & \cellcolor{red!10}0.28
& \cellcolor{red!10}0.48 & \cellcolor{red!10}0.54 & \cellcolor{red!10}0.46 \\
Hyp (8)
& \cellcolor{red!10}0.70 & \cellcolor{red!10}0.75 & \cellcolor{red!10}0.72
& \cellcolor{red!10}0.45 & \cellcolor{red!10}0.61 & \cellcolor{red!10}0.42
& \cellcolor{red!10}0.30 & \cellcolor{red!10}0.26 & \cellcolor{red!10}0.29
& \cellcolor{red!10}0.48 & \cellcolor{red!10}0.54 & \cellcolor{red!10}0.48 \\
\sysname{} (w e5-large-v2)
& \cellcolor{green!15}\textbf{0.88} & \cellcolor{green!15}\textbf{0.99} & \cellcolor{green!15}\textbf{0.85}
& \cellcolor{green!15}\textbf{0.75} & \cellcolor{green!15}\textbf{0.93} & \cellcolor{green!15}\textbf{0.71}
& \cellcolor{green!15}\textbf{0.59} & \cellcolor{green!15}\textbf{0.63} & \cellcolor{green!15}\textbf{0.66}
& \cellcolor{green!15}\textbf{0.74} & \cellcolor{green!15}\textbf{0.85} & \cellcolor{green!15}\textbf{0.74} \\
\sysname{} (w MiniLM)
& \cellcolor{green!15}\textbf{0.93} & \cellcolor{green!15}\textbf{0.99} & \cellcolor{green!15}\textbf{0.92}
& \cellcolor{green!15}\textbf{0.91} & \cellcolor{green!15}\textbf{0.69} & \cellcolor{green!15}\textbf{0.75}
& \cellcolor{green!15}\textbf{0.63} & \cellcolor{green!15}\textbf{0.66} & \cellcolor{green!15}\textbf{0.56}
& \cellcolor{green!15}\textbf{0.82} & \cellcolor{green!15}\textbf{0.78} & \cellcolor{green!15}\textbf{0.74} \\
\sysname{} (w BGE)
& \cellcolor{green!15}\textbf{0.93} & \cellcolor{green!15}\textbf{0.99} & \cellcolor{green!15}\textbf{0.91}
& \cellcolor{green!15}\textbf{0.75} & \cellcolor{green!15}\textbf{0.93} & \cellcolor{green!15}\textbf{0.71}
& \cellcolor{green!15}\textbf{0.63} & \cellcolor{green!15}\textbf{0.66} & \cellcolor{green!15}\textbf{0.57}
& \cellcolor{green!15}\textbf{0.77} & \cellcolor{green!15}\textbf{0.86} & \cellcolor{green!15}\textbf{0.73} \\
\bottomrule
\end{tabular}
}
\end{table*}

Similar pattern holds across other queries. For a query, \texttt{What county is incline village nv?}. Here the keywords \{\texttt{Incline}, \texttt{Nevada}\} identify the location terms that distinguish the relevant document from an irrelevant document with keywords \{\texttt{Airport}, \texttt{Regional}\}. As a result the score of the irrelevant documents drops from 0.41 to 0.11, as the recognized terms don't answer a location query. For the medical query about diabetic bleeding risk, the relevant document surfaces \{\texttt{diabetes}, \texttt{coronary}\} while the irrelevant document shows \{\texttt{behavioural}, \texttt{metabolism}\}. The \sysname{} identify which terms are relevant to the specific medical question.

% The Moore-Penrose pseudoinverse makes this interpretability possible by mapping modulation vectors back to actual vocabulary tokens. Without it, we would only see that scores changed, not which concepts drove those changes.

The method identifies tokens whose embeddings align with the modulation direction; these are the semantic features that the adapters emphasized or suppressed. \texttt{peso} appears in the keyword list, clearly indicating that the modulation moved the embedding toward the dimension where \texttt{peso} lives in the vocabulary space. The ranking changes validate that keywords capture real semantic reasoning. In queries 1 and 3, the correct document doesn't appear in the original top-5 but appears in top-5 after modulation. Here the extracted keywords, \{\texttt{Incline}, \texttt{Nevada}\} and \{\texttt{diabetes}, \texttt{coronary}\}, are the key concepts. In query 2, the relevant document advances from rank 2 to rank 1, with \texttt{peso} as the decisive keyword. This indicates that the system isn't just re-scoring randomly as it identifies relevant semantic concepts for a given query and adjusts the embeddings accordingly. 

Relevant documents show small similarity changes ($\Delta$ = -0.06 to -0.08) because the modulation preserves already-good matches while refining them. Irrelevant documents show large drops ($\Delta$ = -0.29 to -0.34) because the modulation actively suppresses mismatches. This asymmetry demonstrates principled behavior: it strengthens correct alignments and weakens incorrect ones.

% ---------- Table 2 ----------
\begin{table*}[tb]
\centering
\footnotesize
\caption{\textbf{Results on Domain-Specific Datasets (Scifact, Trec-COVID, and Webis-Touche2020).}}
\label{tab:sci-covid-webis}
\resizebox{\textwidth}{!}{%
\begin{tabular}{l|p{0.5cm}p{0.5cm}p{0.5cm}|ccc|ccc|ccc}
\toprule
\textbf{Methods} &
\multicolumn{3}{c|}{\textbf{Scifact}} &
\multicolumn{3}{c|}{\textbf{Trec-COVID}} &
\multicolumn{3}{c|}{\textbf{Webis-Touche}} &
\multicolumn{3}{c}{\textbf{Average}} \\
& nDCG & R & MRR & nDCG & R & MRR & nDCG & R & MRR & nDCG & R & MRR \\
\midrule
e5\mbox{-}large\mbox{-}v2
& \cellcolor{red!10}0.72 & \cellcolor{red!10}0.87 & \cellcolor{red!10}0.68
& \cellcolor{red!10}0.79 & \cellcolor{red!10}0.03 & \cellcolor{green!15}\textbf{1.00}
& \cellcolor{red!10}0.60 & \cellcolor{red!10}0.31 & \cellcolor{red!10}0.85
& \cellcolor{red!10}0.70 & \cellcolor{red!10}0.40 & \cellcolor{red!10}0.84 \\
MiniLM
& \cellcolor{red!10}0.79 & \cellcolor{red!10}0.63 & \cellcolor{red!10}0.66
& \cellcolor{red!10}0.75 & \cellcolor{red!10}0.02 & \cellcolor{red!10}0.87
& \cellcolor{red!10}0.59 & \cellcolor{red!10}0.30 & \cellcolor{red!10}0.85
& \cellcolor{red!10}0.62 & \cellcolor{red!10}0.31 & \cellcolor{red!10}0.65 \\
BGE
& \cellcolor{red!10}0.67 & \cellcolor{red!10}0.84 & \cellcolor{red!10}0.66
& \cellcolor{red!10}0.76 & \cellcolor{red!10}0.02 & \cellcolor{red!10}0.92
& \cellcolor{red!10}0.56 & \cellcolor{red!10}0.29 & \cellcolor{red!10}0.85
& \cellcolor{red!10}0.62 & \cellcolor{red!10}0.38 & \cellcolor{red!10}0.71 \\
DIME 20\%
& \cellcolor{red!10}0.72 & \cellcolor{red!10}0.87 & \cellcolor{red!10}0.67
& \cellcolor{red!10}0.82 & \cellcolor{red!10}0.03 & \cellcolor{green!15}\textbf{1.00}
& \cellcolor{red!10}0.59 & \cellcolor{red!10}0.30 & \cellcolor{red!10}0.85
& \cellcolor{red!10}0.71 & \cellcolor{red!10}0.40 & \cellcolor{red!10}0.84 \\
DIME 40\%
& \cellcolor{red!10}0.70 & \cellcolor{red!10}0.87 & \cellcolor{red!10}0.65
& \cellcolor{red!10}0.82 & \cellcolor{red!10}0.03 & \cellcolor{green!15}\textbf{1.00}
& \cellcolor{red!10}0.61 & \cellcolor{red!10}0.32 & \cellcolor{red!10}0.85
& \cellcolor{red!10}0.71 & \cellcolor{red!10}0.41 & \cellcolor{red!10}0.83 \\
DIME 60\%
& \cellcolor{red!10}0.68 & \cellcolor{red!10}0.84 & \cellcolor{red!10}0.64
& \cellcolor{red!10}0.84 & \cellcolor{red!10}0.03 & \cellcolor{red!10}0.94
& \cellcolor{red!10}0.60 & \cellcolor{red!10}0.31 & \cellcolor{red!10}0.85
& \cellcolor{red!10}0.71 & \cellcolor{red!10}0.39 & \cellcolor{red!10}0.81 \\
DIME 80\%
& \cellcolor{red!10}0.60 & \cellcolor{red!10}0.78 & \cellcolor{red!10}0.55
& \cellcolor{red!10}0.77 & \cellcolor{red!10}0.03 & \cellcolor{red!10}0.86
& \cellcolor{green!15}\textbf{0.62} & \cellcolor{red!10}0.32 & \cellcolor{red!10}0.78
& \cellcolor{red!10}0.66 & \cellcolor{red!10}0.38 & \cellcolor{red!10}0.73 \\
SearchAd
& \cellcolor{red!10}0.69 & \cellcolor{red!10}0.85 & \cellcolor{red!10}0.65
& \cellcolor{red!10}0.84 & \cellcolor{red!10}0.03 & \cellcolor{green!15}\textbf{1.00}
& \cellcolor{red!10}0.58 & \cellcolor{red!10}0.31 & \cellcolor{red!10}0.85
& \cellcolor{red!10}0.70 & \cellcolor{red!10}0.40 & \cellcolor{red!10}0.83 \\
Hyp (2)
& \cellcolor{red!10}0.62 & \cellcolor{red!10}0.75 & \cellcolor{red!10}0.63
& \cellcolor{red!10}0.78 & \cellcolor{red!10}0.02 & \cellcolor{red!10}0.59
& \cellcolor{red!10}0.59 & \cellcolor{red!10}0.29 & \cellcolor{red!10}0.83
& \cellcolor{red!10}0.66 & \cellcolor{red!10}0.35 & \cellcolor{red!10}0.68 \\
Hyp (4)
& \cellcolor{red!10}0.61 & \cellcolor{red!10}0.63 & \cellcolor{red!10}0.68
& \cellcolor{red!10}0.76 & \cellcolor{red!10}0.01 & \cellcolor{red!10}0.44
& \cellcolor{red!10}0.59 & \cellcolor{red!10}0.28 & \cellcolor{red!10}0.83
& \cellcolor{red!10}0.65 & \cellcolor{red!10}0.31 & \cellcolor{red!10}0.65 \\
Hyp (6)
& \cellcolor{red!10}0.72 & \cellcolor{red!10}0.62 & \cellcolor{red!10}0.60
& \cellcolor{red!10}0.79 & \cellcolor{red!10}0.01 & \cellcolor{red!10}0.47
& \cellcolor{red!10}0.60 & \cellcolor{red!10}0.29 & \cellcolor{red!10}0.84
& \cellcolor{red!10}0.70 & \cellcolor{red!10}0.31 & \cellcolor{red!10}0.64 \\
Hyp (8)
& \cellcolor{red!10}0.65 & \cellcolor{red!10}0.81 & \cellcolor{red!10}0.68
& \cellcolor{red!10}0.84 & \cellcolor{red!10}0.01 & \cellcolor{red!10}0.52
& \cellcolor{red!10}0.60 & \cellcolor{red!10}0.29 & \cellcolor{red!10}0.85
& \cellcolor{red!10}0.70 & \cellcolor{red!10}0.37 & \cellcolor{red!10}0.68 \\
\sysname{} (w e5-large-v2)
& \cellcolor{green!15}\textbf{0.74} & \cellcolor{green!15}\textbf{0.88} & \cellcolor{green!15}\textbf{0.70}
& \cellcolor{green!15}\textbf{0.85} & \cellcolor{green!15}\textbf{0.04} & \cellcolor{green!15}\textbf{1.00}
& \cellcolor{green!15}\textbf{0.62} & \cellcolor{green!15}\textbf{0.34} & \cellcolor{green!15}\textbf{0.91}
& \cellcolor{green!15}\textbf{0.74} & \cellcolor{green!15}\textbf{0.42} & \cellcolor{green!15}\textbf{0.87} \\
\sysname{} (w MiniLM)
& \cellcolor{green!15}\textbf{0.82} & \cellcolor{green!15}\textbf{0.65} & \cellcolor{green!15}\textbf{0.69}
& \cellcolor{green!15}\textbf{0.78} & \cellcolor{green!15}\textbf{0.03} & \cellcolor{green!15}\textbf{1.00}
& \cellcolor{green!15}\textbf{0.61} & \cellcolor{green!15}\textbf{0.31} & \cellcolor{green!15}\textbf{0.90}
& \cellcolor{green!15}\textbf{0.65} & \cellcolor{green!15}\textbf{0.35} & \cellcolor{green!15}\textbf{0.68} \\
\sysname{} (w BGE)
& \cellcolor{green!15}\textbf{0.70} & \cellcolor{green!15}\textbf{0.88} & \cellcolor{green!15}\textbf{0.70}
& \cellcolor{green!15}\textbf{0.80} & \cellcolor{green!15}\textbf{0.04} & \cellcolor{green!15}\textbf{1.00}
& \cellcolor{green!15}\textbf{0.60} & \cellcolor{green!15}\textbf{0.31} & \cellcolor{green!15}\textbf{0.90}
& \cellcolor{green!15}\textbf{0.66} & \cellcolor{green!15}\textbf{0.41} & \cellcolor{green!15}\textbf{0.75} \\
\bottomrule
\end{tabular}
}
\end{table*}

% ---------- Table 3 ----------
\begin{table}[tb]
\small
\centering
\caption{\textbf{Results on FiQA.}}
\label{tab:fiqa-only}
\begin{tabular}{l|ccc}
\toprule
\textbf{Methods} & \multicolumn{3}{c}{\textbf{FiQA}} \\
& nDCG & R & MRR \\
\midrule
e5\mbox{-}large\mbox{-}v2
& \cellcolor{red!10}0.20 & \cellcolor{red!10}0.23 & \cellcolor{red!10}0.28 \\
MiniLM
& \cellcolor{red!10}0.22 & \cellcolor{red!10}0.26 & \cellcolor{red!10}0.32 \\
BGE
& \cellcolor{red!10}0.23 & \cellcolor{red!10}0.27 & \cellcolor{red!10}0.34 \\
DIME 20\%
& \cellcolor{red!10}0.20 & \cellcolor{red!10}0.22 & \cellcolor{red!10}0.28 \\
DIME 40\%
& \cellcolor{red!10}0.19 & \cellcolor{red!10}0.21 & \cellcolor{red!10}0.27 \\
DIME 60\%
& \cellcolor{red!10}0.18 & \cellcolor{red!10}0.20 & \cellcolor{red!10}0.25 \\
DIME 80\%
& \cellcolor{red!10}0.14 & \cellcolor{red!10}0.16 & \cellcolor{red!10}0.23 \\
SearchAd
& \cellcolor{red!10}0.17 & \cellcolor{red!10}0.20 & \cellcolor{red!10}0.25 \\
Hyp (2)
& \cellcolor{red!10}0.32 & \cellcolor{red!10}0.40 & \cellcolor{red!10}0.39 \\
Hyp (4)
& \cellcolor{red!10}0.31 & \cellcolor{red!10}0.42 & \cellcolor{red!10}0.35 \\
Hyp (6)
& \cellcolor{red!10}0.32 & \cellcolor{red!10}0.42 & \cellcolor{red!10}0.37 \\
Hyp (8)
& \cellcolor{red!10}0.33 & \cellcolor{red!10}0.42 & \cellcolor{red!10}0.39 \\
\sysname{}(e5\mbox{-}large\mbox{-}v2)
& \cellcolor{red!10}0.22 & \cellcolor{red!10}0.24 & \cellcolor{red!10}0.29 \\
\sysname{} (w MiniLM)
& \cellcolor{red!10}0.23 & \cellcolor{red!10}0.27 & \cellcolor{red!10}0.33 \\
\sysname{} (w BGE)
& \cellcolor{red!10}0.25 & \cellcolor{red!10}0.29 & \cellcolor{red!10}0.37 \\
\sysname{} (Hyp(8))
& \cellcolor{green!15}\textbf{0.36} & \cellcolor{green!15}\textbf{0.44} & \cellcolor{green!15}\textbf{0.42} \\
\bottomrule
\end{tabular}
\end{table}

\section{Discussion on Retrieval Performance}

% ---------- Table 1 ----------

Tables~\ref{tab:ms-nq-hotpot}, \ref{tab:sci-covid-webis}, and \ref{tab:fiqa-only} demonstrate that \sysname{}' bidirectional modulation mechanism consistently improves ranking quality across diverse retrieval scenarios. The performance gains stem from how the adapters reshape the embedding space: the Query Adapter generates transformations that pull semantically relevant documents closer to the query while pushing irrelevant ones away, and the Document Adapter aggregates corpus-level signals to help the query align with the vocabulary and semantic structure of available documents. This dual adaptation addresses the core limitation of static embeddings, they cannot adjust to query-specific relevance signals or corpus-specific terminology. The nDCG rewards systems that place highly relevant documents at top positions with logarithmic discounting~\cite{jarvelin2002cumulated}. By modulating embeddings to strengthen correct semantic alignments, \sysname{} ensures that the most relevant documents rise to ranks 1-3 where users actually look, rather than languishing at ranks 8-10 where they contribute little to user satisfaction.

% The baseline methods' performance reveals why per-query semantic adaptation matters.
\textsc{Search-adaptor} learns dataset-level residual transformations that apply uniformly across all queries, but struggles because different queries require different semantic adjustments, a geographic query about ``incline village'' needs location-term emphasis, while a medical query about ``diabetic bleeding risk'' needs disease-term emphasis. A single global transformation cannot capture this diversity, explaining why \textsc{search-adaptor} often underperforms the base retriever. \textsc{Dime} variants progressively degrade as they remove more dimensions, demonstrating that magnitude-based selection discards information critical for semantic matching, a dimension that appears unimportant globally may be essential for specific queries. \textsc{Hypencoder} generates query-conditioned neural scorers, but these black-box functions lack the explicit semantic grounding that \sysname{}' modulation vectors provide, and adding more layers (2$\rightarrow$8) yields diminishing returns without addressing the fundamental need for interpretable semantic adaptation. The consistent pattern across open-domain datasets (MS MARCO, Natural Questions, HotpotQA) and specialized domains (SciFact, TREC-COVID, FiQA, Webis-Touché) confirms that \sysname{}' approach generalizes: the mechanism adapts to whatever semantic features matter for each dataset, whether lexical overlap, multi-hop reasoning, domain terminology, or argumentative stance.

The FiQA anomaly (Table~\ref{tab:fiqa-only}) illuminates a critical dependency: \sysname{} amplifies the quality of their base embeddings rather than replacing them. When e5-large-v2, MiniLm, and BGE embeddings are poorly calibrated for financial terminologies, \sysname{} built atop them cannot match \textsc{hypencoder}, which uses its own embedding generation. However, stacking \sysname{} on top of Hypencoder (8) embeddings yields substantial gains (9.09\% nDCG, 4.76\% Recall, 7.68\% MRR), demonstrating that the modulation mechanism successfully enhances any sufficiently rich embedding space. This reveals \sysname{} 's architectural advantage: the adapters operate as a plug-and-play layer that improves whatever base retriever provides the best embeddings for a given domain, rather than requiring full model retraining or domain-specific architecture changes. The modulation vectors capture semantic refinements that static embeddings miss, emphasizing currency-related dimensions for financial queries, location dimensions for geographic queries, and disease dimensions for medical queries, while preserving the computational efficiency of cosine similarity.

Comparing base retreivers, MiniLM, BGE and e5 with and without IMRNN, it is clear that with MiniLM-v6, IMRNNs yielded a 3.25\% increase in MRR, 5.12\% in Recall, and 3.64\% in NDCG. These improvements were even more pronounced on BGE, where IMRNNs achieved gains of 5.2\% in MRR, 4.4\% in Recall, and 5.3\% in NDCG. We observe that the magnitude of improvement scales with the model's complexity as evident from comparing MiniLM, BGE and e5.

\section{Conclusion}

We introduce \sysname{}, the first lightweight retrieval adapters that make the embeddings of dense retrievers interpretable by achieving three levels of interpretability: structural, attribution, and semantic. We benchmark the retrieval performance of \sysname{} against state-of-the-art retrieval adapter baselines on diverse datasets and demonstrate that \sysname{} adapts query and document embeddings more effectively than competing methods, while also showing strong generalization. The semantic-level interpretability of \sysname{} is especially useful in applications where access to key tokens or keywords in both the query and the documents plays a major role. %Lastly, \sysname{} presents an interpretable workflow of the RAG system, and make it 
%is a step toward making key architectures such as RAG more 
We release all code and scripts under the CC BY 4.0 license for reproducibility.\footnote{\url{https://github.com/YashSaxena21/IMRNNs}}

% \newpage

\section{Limitations}
Experiments reveal three key limitations of \sysname{}. First, the token-level attribution method can produce noisy mappings where some identified tokens (e.g., ``Ridges'', ``Innate'' in Table \ref{tab:qualitative_analysis}) appear semantically unclear or spuriously correlated with the query-document relationship. This occurs because back-projecting continuous embedding modulations to discrete tokens via pseudoinverse is inherently approximate, and the closest token in embedding space may not semantically correspond to the actual concept driving the modulation. Filtering is often necessary to obtain interpretable explanations, but systematic methods for identifying spurious tokens remain an open challenge. Second, \sysname{} incur higher inference latency than dimension-selection methods (Table \ref{tab:inference_latency_throughput} in Appendix) because the Document Adapter must process each corpus document individually to compute transformations. While this bidirectional modulation enables richer semantic adaptation, it scales linearly with corpus size, potentially limiting deployment on extremely large corpora ($>$10M documents) without infrastructure optimizations like caching or approximate nearest neighbor filtering. Third, IMRNNs amplify rather than replace the quality of base embeddings. When base retrievers produce poorly calibrated embeddings for a domain (e.g., financial terminology in FiQA), IMRNNs cannot compensate for fundamental semantic gaps. The modulation mechanism assumes the base embedding space already captures relevant semantic dimensions, it refines their emphasis rather than introducing new concepts. This dependency suggests IMRNNs are best deployed as an enhancement layer atop domain-appropriate base retrievers rather than a universal solution.

\section*{Acknowledgements}
We thank the ACL ARR reviewers for their constructive feedback that significantly improved this work. We are grateful to Mandar Chaudhary and the students in the Knowledge-infused AI and Inference Lab at UMBC for their insightful discussions and reviews. This work was supported in part by USISTEF and the UMBC Cybersecurity Initiative. The views and conclusions contained herein are those of the authors and should not be interpreted as representing the official policies of the funding agencies.

% \bibliographystyle{ACM-Reference-Format}
% \newpage
\bibliography{main}
\appendix

\section{Dataset Selection}
\label{sec:Dataset_selection}

BEIR contains 15 retrieval datasets that cover a wide range of domains and query types. In the main paper, we selected seven datasets that provide sufficient diversity for evaluating both retrieval effectiveness and the interpretability mechanisms introduced by \sysname{}. The goal was to include datasets that differ meaningfully in domain, reasoning structure, and retrieval difficulty, rather than to exhaustively evaluate on all BEIR tasks.

Our chosen subset covers three important axes of variation:

\begin{itemize}
    \item \textbf{Domain diversity.} The selected datasets span open-domain retrieval (MS MARCO, NQ), multi-hop reasoning (HotpotQA), scientific fact checking (Scifact), legal and argument-focused retrieval (Webis-Touche), biomedical retrieval (Trec-COVID), and financial question answering (FiQA). This includes all major retrieval settings that stress different aspects of semantic matching.

    \item \textbf{Query complexity.} We include datasets that require single-hop retrieval, multi-hop synthesis, and fact verification. This variation is important because the modulation mechanism in \sysname{} adapts embeddings based on both query semantics and corpus structure, which is not tested by simple single-hop retrieval alone.

    \item \textbf{Dataset size and structural properties.} The selected datasets cover a wide range of corpus sizes and document distributions. This ensures that \sysname{} is evaluated under different levels of retrieval sparsity, redundancy, and noise. These conditions affect how much semantic refinement the adapters can provide.
\end{itemize}

Several of the remaining BEIR datasets are stylistic variants of tasks already included or add little new retrieval structure relative to the selected subset. This selection approach is consistent with many recent retrieval studies that report results on a representative subset of BEIR when the full benchmark is not required to evaluate the proposed contribution. Our method is dataset-agnostic because interpretability in \sysname{} arises from the mathematical properties of modulation and back-projection, rather than dataset-specific lexical patterns. Adding more datasets would increase volume but would not change the qualitative insights or the interpretability analysis.

\begin{table}[!t]
\centering
\small
\caption{\textbf{BEIR Dataset Details.}}
\label{app:dataset-details}
\resizebox{0.49\textwidth}{!}{%
\begin{tabular}{l l c r r}
\toprule
\textbf{Dataset} & \textbf{Domain} & \textbf{Type} & \textbf{\#Queries} & \textbf{\#Documents} \\
\midrule
MS MARCO       & Web Search      & Single-hop & 6{,}980  & 8{,}841{,}823 \\
Natural Questions & Wikipedia    & Single-hop & 3{,}452  & 2{,}681{,}468 \\
HotpotQA       & Wikipedia       & Multi-hop  & 7{,}405  & 5{,}233{,}329 \\
SciFact        & Scientific      & Single-hop & 300      & 5{,}183 \\
TREC-COVID     & Biomedical      & Single-hop & 50       & 171{,}332 \\
Webis-Touché 2020 & Argumentation & Single-hop & 49       & 382{,}545 \\
FiQA-2018      & Finance         & Single-hop & 648      & 57{,}638 \\
\bottomrule
\end{tabular}
}
\end{table}

\section{Implementation Details}
\label{app:impl-details}
 \textsc{IMRNNs} use projection dimension $m=256$. Both adapters are two-layer MLPs with ReLU activations and layer normalization. Training uses Adam (learning rate $10^{-4}$, weight decay $10^{-5}$, batch size 32) with margin $\gamma=0.3$, operating on top-100 BM25 candidates per query. Early stopping uses patience of 5 epochs on validation nDCG. Training converges within 10-20 epochs on a single NVIDIA H100 GPU.

 \section{Computational Time Comparison}
 \label{sec:compute-time}

Latency is averaged per single query, and throughput is computed as the inverse of average latency.

 \begin{table}[!htbp]
\centering
\caption{\footnotesize Inference efficiency comparison (averaged over 1{,}000 queries).}
\label{tab:inference_latency_throughput}
\setlength{\tabcolsep}{8pt}
\renewcommand{\arraystretch}{0.95}
\resizebox{\columnwidth}{!}{
\begin{tabular}{lcc}
\toprule
\textbf{Method} & \textbf{Latency (ms/query)} & \textbf{Throughput (queries/s)} \\
\midrule
Magnitude DIME & 0.96 & 61.22 \\
Search-Adaptor & 1.02 & 58.82 \\
Hypencoder & 1.71 & 35.09 \\
\sysname{} & 1.64 & 36.59 \\
\bottomrule
\end{tabular}
}
\vspace{-3pt}
\begin{flushleft}
% \footnotesize \emph{Notes.} Latency is averaged per single query; throughput is computed as inverse of average latency.
\end{flushleft}
\end{table}

\section{Additional Experiments}

% ---------- Table: MiniLM Open-Domain ----------
\begin{table*}[tb]
\centering
\small
\caption{\textbf{Additional Experiments on ArguAna, Quora, and Scidocs datasets from the BEIR benchmark suite.}}
\label{tab:minilm_open}
\resizebox{\textwidth}{!}{%
\begin{tabular}{l|ccc|ccc|ccc|ccc}
\toprule
\textbf{Methods} &
\multicolumn{3}{c|}{\textbf{ArguAna}} &
\multicolumn{3}{c|}{\textbf{Quora}} &
\multicolumn{3}{c|}{\textbf{Scidocs}} &
\multicolumn{3}{c}{\textbf{Average}} \\
& nDCG & R & MRR & nDCG & R & MRR & nDCG & R & MRR & nDCG & R & MRR \\
\midrule
e5-large
& \cellcolor{red!10}0.78 & \cellcolor{red!10}0.94 & \cellcolor{red!10}0.87
& \cellcolor{red!10}0.90 & \cellcolor{red!10}0.89 & \cellcolor{red!10}0.96
& \cellcolor{red!10}0.25 & \cellcolor{red!10}0.29 & \cellcolor{red!10}0.95
& \cellcolor{red!10}0.64 & \cellcolor{red!10}0.70 & \cellcolor{red!10}0.92 \\
MiniLM
& \cellcolor{red!10}0.86 & \cellcolor{red!10}0.72 & \cellcolor{red!10}0.75
& \cellcolor{red!10}0.88 & \cellcolor{red!10}0.84 & \cellcolor{red!10}0.92
& \cellcolor{red!10}0.15 & \cellcolor{red!10}0.22 & \cellcolor{red!10}0.89
& \cellcolor{red!10}0.63 & \cellcolor{red!10}0.59 & \cellcolor{red!10}0.85 \\
BGE
& \cellcolor{red!10}0.75 & \cellcolor{red!10}0.92 & \cellcolor{red!10}0.88
& \cellcolor{red!10}0.89 & \cellcolor{red!10}0.84 & \cellcolor{red!10}0.90
& \cellcolor{red!10}0.20 & \cellcolor{red!10}0.27 & \cellcolor{red!10}0.92
& \cellcolor{red!10}0.61 & \cellcolor{red!10}0.67 & \cellcolor{red!10}0.90 \\
\sysname{} (e5-large)
& \cellcolor{green!15}\textbf{0.81} & \cellcolor{green!15}\textbf{0.96} & \cellcolor{green!15}\textbf{0.92}
& \cellcolor{green!15}\textbf{0.91} & \cellcolor{green!15}\textbf{0.92} & \cellcolor{green!15}\textbf{0.98}
& \cellcolor{green!15}\textbf{0.29} & \cellcolor{green!15}\textbf{0.34} & \cellcolor{green!15}\textbf{0.96}
& \cellcolor{green!15}\textbf{0.67} & \cellcolor{green!15}\textbf{0.74} & \cellcolor{green!15}\textbf{0.95} \\
\sysname{} (MiniLM)
& \cellcolor{green!15}\textbf{0.90} & \cellcolor{green!15}\textbf{0.76} & \cellcolor{green!15}\textbf{0.79}
& \cellcolor{green!15}\textbf{0.91} & \cellcolor{green!15}\textbf{0.92} & \cellcolor{green!15}\textbf{0.99}
& \cellcolor{green!15}\textbf{0.19} & \cellcolor{green!15}\textbf{0.25} & \cellcolor{green!15}\textbf{0.91}
& \cellcolor{green!15}\textbf{0.66} & \cellcolor{green!15}\textbf{0.64} & \cellcolor{green!15}\textbf{0.89} \\
\sysname{} (BGE)
& \cellcolor{green!15}\textbf{0.79} & \cellcolor{green!15}\textbf{0.96} & \cellcolor{green!15}\textbf{0.90}
& \cellcolor{green!15}\textbf{0.92} & \cellcolor{green!15}\textbf{0.89} & \cellcolor{green!15}\textbf{0.94}
& \cellcolor{green!15}\textbf{0.27} & \cellcolor{green!15}\textbf{0.29} & \cellcolor{green!15}\textbf{0.95}
& \cellcolor{green!15}\textbf{0.66} & \cellcolor{green!15}\textbf{0.71} & \cellcolor{green!15}\textbf{0.93} \\
\bottomrule
\end{tabular}
}
\end{table*}

The additional experiments provided in \autoref{tab:minilm_open} strengthen the main claim that \sysname{} is retriever-agnostic at the same time dataset independent and can be attached to a wide range of dense encoders without retraining them. Two observations are consistent across all models.

\begin{itemize}
    \item The magnitude of improvement grows with the capacity of the base retriever. MiniLM performs the worst among the three base models, and its gains are smaller, while BGE performs better and shows larger improvements.
    \item Even compact models benefit from modulation. The adapters consistently improve the separation between relevant and irrelevant documents regardless of the dimensionality of the embedding space.
\end{itemize}

Overall, the results provide strong evidence that \sysname{} generalizes well beyond the specific encoder used in the main paper, and can be reliably deployed across diverse retrieval systems.
\end{document}